\documentclass[draftcls,onecolumn,12pt]{IEEEtran}
\usepackage[cmex10]{amsmath}
\usepackage{geometry}                
\usepackage{graphicx}
\usepackage{amssymb}
\usepackage{color}
\usepackage{epstopdf}
\usepackage{amsthm}
\newtheorem{theorem}{Theorem}
\newtheorem{example}{Example}
\newtheorem{definition}{Definition}
\newtheorem{construction}{Construction}

\makeatletter

\begin{document}
\title{\Large  \sc Query Matrices for Retrieving Binary Vectors Based on the Hamming Distance Oracle}
\author{{Vinay A. Vaishampayan}\\
{AT\&T Labs-Research,
Shannon Laboratory\\
180 Park Avenue, Florham Park, NJ 07932\\
{\tt vinay@research.att.com}
}}

\maketitle
\date{\today}
\subsection*{\centering Abstract}
{\em
The Hamming oracle returns the Hamming distance  between an unknown binary  $n$-vector $x$ and a binary query $n$-vector y.
The objective is to determine $x$ uniquely using a sequence of $m$ queries.
What are the minimum number of queries required in the worst case?
We consider the query ratio $m/n$ to be our figure of merit and derive
upper bounds on the query ratio  by explicitly constructing
$(m,n)$ query matrices. We show that
our recursive and algebraic construction results in query ratios arbitrarily close to zero. Our construction is
based on codes of constant weight. A decoding algorithm for recovering the unknown binary vector is
also described.
}
\section{Introduction}
\label{sec-intro}
An unknown binary vector $x$ of length $n$ bits is to be determined by posing a sequence of queries to an oracle. Each query takes the form of a binary vector $y$ of length $n$ bits.  We consider two kinds of oracles, (i) the Hamming distance oracle that returns the Hamming distance $d(x,y)$ between $x$ and $y$, and (ii) the overlap oracle that returns the Hamming weight $w(x.y)$ of the inner product between $x$ and  $y$. Our measure of efficiency is the number of queries required to determine $x$ for the worst $x$ and our objective is to determine a sequence of queries that minimizes the number of queries in the worst case. There are two cases of interest: (i) {\it adaptive}: the $i$th query is allowed to depend on all preceding queries and responses, and (ii) {\it non-adaptive}:  the queries are formulated in advance. We only consider the non-adaptive case in this paper. 

The binary set $\{0,1\}$ is treated as a subset of the integers. The weight of a binary vector, denoted $w(x)$ counts the number of ones in $x$ and the Hamming distance $d(x,y)$  counts the number of positions in which $x$ and $y$ differ.
The inner product is the standard Euclidean inner product $\sum_{i=1}^nx_i y_i$.  

The overlap oracle  is  related to the group testing oracle~\cite{DuHwang:1993}. In group testing, an unknown binary $n$-vector $x$ has 
Hamming weight $w(x) \leq d$ for some given $d \leq n$. A query $y$ is a binary $n$-vector  and the oracle
returns  $1$ if $w(x.y)>0$ and $0$ if $w(x.y)=0$.  The main difference between the problem studied here and the group testing problem is that (i) the group testing oracle is less informative, and (ii) we impose no restriction on the Hamming weight of the
unknown vector $x$. 
Since the Hamming   and overlap oracles are more informative than the group testing oracle, and  we expect that fewer queries will suffice in order to determine $x$.
Previous work on the Hamming oracle includes~\cite{Ewert:2010} and \cite{Maurer:web} where it is observed that $m < n$ queries suffice to determine $x$.  The problem is closely related
to the distinct subset sum problem, which is the problem of constructing sets of natural numbers such that the sum over any subset is unique. The reader is referred to
papers by Conway and Guy~\cite{ConwayGuy:1968}, Guy~\cite{Guy:1982}, Lunnon~\cite{Lunnon:1988} and  Bohman~\cite{Bohman:1996} among others. 

The paper is organized as follows. Basic notation, and some examples are presented in Sec~\ref{sec-prelim}. A lower bound
on the query ratio is proved in Sec.~\ref{sec-gen} using a packing argument. Also a previously known upper bound~\cite{LevYuster:2011}, based on a probabilistic argument is stated. The discrete subset sum problem is described in Sec.~\ref{sec-dss}, and a preliminary construction is given. Sec.~\ref{sec-back} contains some relevant results on constant weight codes and block designs. Our basic construction is given in Sec.~\ref{sec-alg-1}, followed by an iterated construction in Sec.~\ref{sec-alg-2}. A decoding method is described in Sec.~\ref{sec-decoding}, achievable query ratios are derived in Sec.~\ref{sec-achievable}, two example constructions are given in Sec~\ref{sec-examples} and conclusions are in Sec.~\ref{sec-conclusions}.

\section{Preliminaries}
\label{sec-prelim}
All vectors are column vectors.   From the identity $d(x,y)=w(x)+w(y)-2w(x.y)$, it follows that if $w(x)$ is known, then $w(x.y)$ can be determined from $d(x,y)$. The cost of determining $w(x)$ is a single query---the all-ones query. Thus, if $m$ queries to the overlap oracle are sufficient to determine $x$, then  at most one extra query to a Hamming oracle would suffice to determine $x$. 

For the overlap oracle,  $x$ can be obtained as   a solution to a system of linear equations
\begin{equation}
Qx=\omega
\label{eqn:query}
\end{equation}
where $Q$ is an $(m,n)$ binary  matrix the $i$th row of which is the $i$th query vector, $x$ is the unknown binary vector, $\omega=(\omega_i)$ is an $m$-vector of non-negative integers, $\omega_i=w(x.q_i)$, where $q_i$ is the $i$th row of $Q$ and
multiplication is over the reals. 

If (\ref{eqn:query}) has a unique solution for every $x$, then any non-zero vector in the null space of $Q$, ${\mathcal N}(Q)$ cannot consist solely of entries from the
set $\{-1,0,1\}$ for if it did, then two binary vectors $x_1$ and $x_2$ would satisfy $Q x_1=Q x_2$~\cite{Bohman:1996}, \cite{LevYuster:2011}. This leads to the following definition.
\begin{definition}
Query matrix $Q$ is said to be uniquely identifying (UI) if the only $\{-1,0,1\}$-valued vector in  ${\mathcal N}(Q)$ is the all-zero vector. 
Equivalently,
$Q$ is UI if every disjoint pair of  column subsets have unequal column sums.
\end{definition}
\begin{definition}
Query ratio $\rho$ to said to be achievable if  there exists a UI query matrix $Q$ of size $(m,n)$ with $m/n \leq \rho$. 
\end{definition}
Clearly if $Q=I$ where $I$ is the identity matrix, then $Q$ is UI. Thus $\rho \leq 1$.  It is interesting that  $\rho $ arbitrarily close to $0$ is  achievable. 
\begin{example}
We claim that the $(4,5)$ matix
\begin{equation}
Q=\begin{pmatrix}1 1 1 1 1\\ 0 1 0 0 1\\0 0 1 0 1\\0 0 0 1 1  \end{pmatrix}
\end{equation}  
is UI. Thus $\rho \leq 0.8$.
\label{ex:1}
\end{example}
The proof is accomplished by showing that there is no non-zero vector $z$ with entries from $\{{-1,0,1}\}$ which lies in ${\mathcal N}(Q)$. A non-zero $\{-1,0,1\}$-valued vector $z$ in ${\mathcal N}(Q)$ identifies two disjoint subsets of columns that have identical sums, and in this case the size of each subset must be the same (since the top row is all ones). A subset cannot consist of a single column since all columns are distinct; if a subset consists of two columns, it cannot pair the last column with any other than the first column else one of the lower three positions would be two and cannot equal the sum of any of the remaining two columns. By a similar argument the first column must be paired with the last column. But then, with the first and last columns paired together, 
equal column sums cannot be achieved using any two of the middle three columns for position will always be zero.

In addition to proving that $Q$ is UI, the above example also highlights the difficulties involved in scaling the case analysis to larger matrices. Clearly a more efficient method of proof is required. Also, note that the query matrix in the above example is valid for
both the Hamming and overlap oracles.

For query matrices with an all 1's top row we can show that  a $3\times 4$ matrix does not exist, thus for $n=4$, $m=3$ queries are never
enough for the Hamming oracle. However, three queries suffice for the overlap oracle as the following example shows.
\begin{example}
The  $(3,4)$ matrix $Q=\begin{pmatrix} 1101\\1011\\0111\end{pmatrix}$ is UI.
\end{example}
If we append an all $0$ column to this $Q$ and then place an all $1$'s row on top we get
a $4\times 5$ uniquely identifying matrix for the Hamming oracle.

\section{General Observations}
\label{sec-gen}
Using a  packing argument we prove the following lower bound on query ratio $\rho$ for a UI matrix.
\begin{theorem}
For the Hamming oracle the query ratio must satisfy $\rho=m/n > 1/log_2(n+1)$ for any sequence of $m$ queries that determine every $n$-bit $x$ uniquely.
\label{thm:packing}
\end{theorem}
\begin{IEEEproof}
For each $x$ the Hamming oracle must result in a unique
column $m$-vector of Hamming distances in order to guarantee that $x$ be determined. But there are at most $(n+1)^m$ distinct Hamming distance vectors. Thus $2^n \leq (n+1)^m$ or $\rho=m/n \geq 1/\log_2(n+1)$. 
\end{IEEEproof}

In a recent contribution~\cite{LevYuster:2011}, it has been shown using the probabilistic method, that any 
\begin{equation}
\rho \geq (\log_2 9/\log_2 n) (1+\phi(n))
\label{eqn:LVbound}
\end{equation}
is achievable, where $\phi(n) \rightarrow 0$ as $n \rightarrow \infty$. This was accomplished by showing that binary matrices with $m$ rows and $n$ columns exist, for which no disjoint sets of columns have equal sums.

\section{Relation to the Discrete-Subset-Sum Problem and a Construction}
\label{sec-dss}
A subset of the positive integers, all of whose sums are distinct, is said to have distinct subset sums. We call such a set with $n$ elements, an $n-$DSS set. A well known example of a DSS set is $\{1,2,4,\ldots,2^{n-1}\}$.  If we denote the $i$th element of $S$ by $r_i$,  then for the exponential set $r_n=(1/2) 2^n$. The objective is to construct $n$-DSS sets  for which $r(n)$ is small. The current record holding  construction for an $n$-DSS set~\cite{Bohman:1996} achieves  $r_n< 0.22002 \cdot 2^n$. 

The similarity of our problem to the $n$-DSS set problem comes from the observation that for an $n$-DSS set, the only $\{-1,0,1\}$-valued vector $y$ for which $\sum_{i=1}^n r_i y_i
=0$ is the all zero vector $y=0$.

 We now present an elementary construction based on an $n$-DSS set. 
 \begin{construction}
Given an $n$-DSS set $S$, set $m=\lceil \log_2 r(n) \rceil$ and construct the $(m, n)$ matrix $Q$ by setting its $i$th column to 
the $m$-bit binary expansion of $r(i)$, the $i$th element of $S$.  
\end{construction}

\begin{theorem}
The matrix $Q$ of Construction 1 is UI.
\end{theorem}
\begin{IEEEproof}
Suppose the column sums over $\mathcal I$ and $\mathcal J$, two  non-overlapping subsets of columns of $Q$, are identical, and denote these column sums by $x$ and $y$, respectively, where $x$, $y$ are integer vectors of length $\lceil \log_2 r(n)\rceil$. This means $\sum_{i=0}^{m-1}x_i 2^{i}=\sum_{i=0}^{m-1}y_i2^{i-1}$. But $\sum_{i=0}^{m-1} x_i 2^{i}= \sum_{j \in \mathcal I}r(j)$ and this implies $\sum_{j \in \mathcal I}r(j)=\sum_{j \in \mathcal J}r(j)$, a contradiction.  \end{IEEEproof}

Construction 1 shows that  UI query matrices of size $(n-2,n)$ exist for $n$ suitably large.

By reversing the steps of the construction, we can build a subset of the positive integers starting with an $(m,n)$ binary matrix $Q$. However it is not necessary that the resulting set $S$ have discrete set sums. This can be seen with the matrix $Q$ shown in Example~\ref{ex:1}. The resulting set $S=\{1,3,5,9,15\}$ clearly does not have distinct subset sums, even though $Q$ has distinct column sums.

A better method for using DSS sets for the Hamming oracle problem is proposed later in this paper.

\section{Background for Algebraic Constructions}
\label{sec-back}
Material in this Section is drawn from books by MacWilliams and Sloane~\cite{MacwilliamsSloane:1977}, Hall~\cite{Hall:1986}, and Cameron~\cite{Cameron:1994}. 
\cite{Hall:1986}.
\begin{definition}[Hall]
A balanced block design $(b,v,r,k,\lambda)$ is an arrangement of $v$  distinct objects into $b$ blocks such that each block contains exactly $k$ distinct objects, each object occurs in exactly $r$ different blocks, and every pair of distinct objects occurs together in exactly $\lambda$ blocks.
\end{definition}
Balanced block designs are categorized as {\it complete}, which is the set $C^k_v$, all combinations of $k$ objects from a set of $v$ objects, and {\it incomplete}, a proper subset of 
$C^k_v$ in which each pair of objects occurs an equal number of times. 

It is known that every balanced block design must satisfy the identities
\begin{eqnarray}
bk & = & vr \\
r(k-1)& = & \lambda (v-1).
\end{eqnarray}
The following is known about the existence of block designs.
\begin{theorem}[Wilson~\cite{Wilson:1972}]
Given $\lambda$ and $k$, there is a $v_0$, such that if $v \geq v_0$ and $\lambda(v-1) \equiv 0 \pmod{(k-1)}$ and $\lambda v (v-1) \equiv 0 \pmod{k(k-1)}$, then there exists a design $(b,v,r,k,\lambda)$ with $r=\lambda(v-1)/(k-1)$ and $b=\lambda v (v-1)/k(k-1)$.
\end{theorem}
From Wilson's theorem it follows that for $v$ large enough  there exists a $\lambda=1$ design with
\begin{equation}
b=\frac{v(v-1)}{k(k-1)}
\label{eqn:Wbound}
\end{equation}
blocks.
The incidence matrix of a block design is a matrix with $b$ rows and $v$ columns, the $(i,j)$ entry of which is $1$ if block $i$ contains object $j$, and $0$ otherwise. 
In our constructions we will use subsets of such incidence matrices for block designs with $\lambda=1$. Note that rows of the incidence matrix of a pairwise balanced block design 
form a code of constant Hamming weight $k$, block length $v$ and minimum Hamming distance
$d_{min}=2(k-1)$.

Of particular importance is the construction and lower bound given by Graham and Sloane~(Thm. 4,~\cite{GrahamSloane:1980} )
which states that for  $q$ a prime power, $q \geq n$, there exists a constant weight code of weight $w$ and minimum distance $2\delta$ with
\begin{equation}
A(n,2\delta,w) \geq \frac{1}{q^{\delta-1}}{ n \choose w}
\label{eqn:GSbound}
\end{equation}
codewords. Further, it is shown in Th. 6 of \cite{GrahamSloane:1980} that $q$ need not be much greater than $n$.

\section{A Level-1 Algebraic Construction}
\label{sec-alg-1}
Let $n=m+k$ with $k < m$. Thus $m/n> 1/2$. 
Let 
\begin{equation}
Q_1=\begin{pmatrix}
I_k & C_1 & E_1 \\
0 & I_{m-k} & C_1^{tr}
\label{eqn:defA}
\end{pmatrix},
\end{equation}
where $I_k$ is the identity matrix of size $k$, the zero sub-matrix $0$ is  of size $(m-k,k)$, sub-matrix $C_1$ is of size $(k,m-k)$, $E_1$ is of size $(k,k)$ and 
$C_1^{tr}$ denotes the transpose of $C_1$.
For the purposes of this paper we define a specific version  $Q_1(r)$ with $C_1$ chosen as an $({r \choose 2},r)$ matrix, whose rows are all the distinct binary $r$-tuples of Hamming  weight $2$. Thus $k={r \choose 2}$ and  $m-k=r$. The  sub-matrix $E_1$  is given by  $C_1C_1^{tr}-2I=E_1$ (the size of the identity matrix is clear from the context and is not stated explicitly) and has entries from the set $\{0,1\}$ with zeros along the main diagonal. This follows directly from the fact that the inner product of two rows of $C$ is in $\{0,1\}$.

\begin{theorem}
$Q_1(r)$ is UI and achieves a query ratio arbitrarily close to $1/2$ for $r$ suitably large.
\end{theorem}
\begin{IEEEproof}
We show that $Q_1(r)$ does not have a nonzero $\{-1,0,1\}$-valued vector in its null space. Let
\begin{equation}
Q_1 (r)\begin{pmatrix} x \\ y \\ z \end{pmatrix}= \begin{pmatrix}
I_k & C_1 & E_1 \\
0 & I_{m-k} & C_1^{tr}
\end{pmatrix} \begin{pmatrix} x \\ y \\ z \end{pmatrix} =0,
\end{equation}
where $x$, $y$ and $z$ are of size $k$, $m-k$ and $k$, respectively. Thus $y=-C_1^{tr} z$ and $x=(C_1 C_1^{tr} -E_1)z$.
But since $C_1 C_1^{tr} -E_1=2I$  the 
only $\{-1,0,1\}$-valued vector in $\mathcal{N}(Q_1)$ is the zero vector. Thus $Q_1(r)$ is UI.

The query ratio for $Q_1(r)$ is $\rho=\frac{({r \choose 2}+r)}{(2{r \choose 2}+r)}$ which goes to $1/2$ as $r$ grows without bound.
\end{IEEEproof}

Thus to
achieve a ratio $\rho = 0.51$ requires  $r\geq 50$, or equivalently $n \geq  2500$. We can thus 
query $2500$ bits with $1275$ queries using $Q_1(50)$. 

For later use we have the following result.
\begin{theorem}
For any nonzero $n$-vector $x$ with entries in $\{-1,0,1\}$,   $Q_1(r) x $ does not lie in $(4 \mathbb{Z})^m $.
\label{thm:base}
\end{theorem}
\begin{IEEEproof}
Let $x$ be partitioned into sub-vectors $x_1$, $x_2$ and $x_3$ of size $k$, $r$ and $k$ respectively, i.e. let $x=(x_1, x_2, x_3)$. For some  integer $m$-vector $u$, let $Q_1(r)x=4u$. Let $u=(u_1,u_2)$ where $u_1$ and $u_2$, of size $k$ and $m-k$, respectively.
Thus $x_1+C_1 x_2 +E_1 x_3=4u_1$ and $x_2+C^{tr}_1 x_3=4u_2$. Simplification gives us
\begin{equation}
x_1=(C_1 C_1^{tr} -E_1) x_3 +4(u_1-C_1 u_2).
\label{eqn:p3}
\end{equation}
Since $x_1$ takes values in $\{-1,0,1\}$ and since the first term on the right hand side is $2x_3$, the only possible solution for (\ref{eqn:p3})
is $(u_1-C_1u_2)=0$, $x_1=0$ and $x_3=0$. But this means $x_2=4u_2$, the only solution of which is $x_2=0$ and $u_2=0$.
This implies $u_1=0$. Thus $x=0$ and $u=0$ is the only solution for $Q_1(r) x=4u$.
\end{IEEEproof}
\section{An Iterated Higher-Level Construction}
\label{sec-alg-2}
Let $(m_s,n_s)$ be the size of $Q_s(r)$.
\begin{definition}
For $s>1$ we define recursively
\begin{equation}
Q_s(r)=
\begin{pmatrix}
Q_{s-1}(r) & C_s & E_s \\
0 & I & C_s^{tr}
\end{pmatrix},
\label{eqn:Qiter}
\end{equation}
where $C_s$ is a binary  matrix whose rows are of constant Hamming weight $2^s$, the number of rows of $C_s$ is equal to the number of rows of $Q_{s-1}$, and the number of columns of $C_s$ (the block length of the code) is the minimum possible such that  $d_{min}$ the minimum Hamming distance between the rows of $C_s$ is $2(2^s-1)$. $E_s$ is the binary matrix given by  $C_s C_s^{tr} -2^s I =E_s$.
\end{definition}
We now have
\begin{theorem}
For any nonzero $n$-vector $x$ with entries in $\{-1,0,1\}$,   and any $s>0$, $Q_s(r) x $ does not lie in $(2^{s+1} \mathbb{Z})^m $, where $m$ is the number of rows of $Q_s(r)$.
\label{thm:notpower}
\end{theorem}
\begin{IEEEproof}
We have already proved the case $s=1$. We proceed by induction. Suppose the hypothesis holds for $Q_{s-1}$. Consider the equation $Q_{s}(r)x=2^{s+1}u$, where $u$ is an integer-valued vector of appropriate dimension. Following the proof of Thm.~\ref{thm:base}, we
have
\begin{eqnarray}
Q_{s-1}x_1 & = & 2^s x_3 +2^{s+1}(u_1-C_s u_2), \label{eqn:ind-1}\\
x_2 & = &  -C_s^{tr} x_3+2^{s+1}u_2. \label{eqn:ind-2}
\end{eqnarray}
From (\ref{eqn:ind-1}) and the induction hypothesis, it follows that $(u_1-C_su_2)=0$, $x_3=0$ and $x_1=0$. Since $x_3=0$, it follows from (\ref{eqn:ind-2})
that $u_2=0$ and $x_2=0$. Since $u_1-C_s u_2=0$ it follows that $u_1=0$.
\end{IEEEproof}
This leads to our main result.
\begin{theorem}
$Q_s(r)$ is UI for all $s>0$.
\end{theorem}
\begin{IEEEproof}
Proof is by induction. We have already proved that $Q_1(r)$ is UI. Assume the hypothesis is true
for $Q_{s-1}(r)$. Consider the equation
\begin{equation}
Q_{s}(r) \begin{pmatrix} x \\ y \\ z \end{pmatrix}=0
\end{equation}
which can be simplified to 
\begin{eqnarray}
Q_{s-1}(r) x & = &  2^s z  \label{eqn:thm7-1}\\
y & = & -C_{s}^{tr} z \label{eqn:thm7-2}
\end{eqnarray}
using (\ref{eqn:Qiter}). From Thm.~\ref{thm:notpower}, it follows that $z=0$ in
(\ref{eqn:thm7-1}) and since $Q_{s-1}$ is UI, it follows that $x=0$. From (\ref{eqn:thm7-2})
it follows that $y=0$. Thus the only $\{-1,0,1\}$-valued vector in $\mathcal{N}(Q_s)$ is the
zero vector.
\end{IEEEproof}

\section{Decoding Rule}
\label{sec-decoding} For $j=0,1,\ldots,s$, let level-$j$ query matrix $Q_j$ be of size $(m_j,n_j)$ and define $Q_0=I_{m_0}$.
For ease of reference the recursive structure of the query matrix $Q_j$, $j>1$ is repeated here. We define $Q_0(r)$ to be the identity
matrix of size $m_0$.
\begin{equation}
Q_j(r)=
\begin{pmatrix}
Q_{j-1}(r) & C_j & E_j \\
0 & I & C_j^{tr}
\end{pmatrix}.
\label{eqn:Qiter-rep}
\end{equation}
Note that $n_j=n_{j-1} + m_j$, $j=1,2,\ldots,s$.

Given $s$ and the $m_s$-vector of non-negative integers $\omega^{(s)}$ returned by the oracle, our objective
is to solve 
\begin{equation}
Q_s x^{(s)}=\omega^{(s)}
\label{eqn:root}
\end{equation}
for $x^{(s)}$ in $\{0,1\}^{n_s}$. 

It is convenient to write $x^{(j)}=(x^{(j-1)}, y^{(j-1)},z^{(j-1)})$, $j=1,2,\ldots,s$, where
the part  $x^{(j-1)}$ has size $n_{j-1}$, $y^{(j-1)}$ has size $m_j-m_{j-1}$ and $z^{(j-1)}$ has size $m_{j-1}$.
For $a=m_{j-1}$, $b=m_j-m_{j-1}$ and $j=1,2,\ldots,s$ we have the following definitions:
\begin{eqnarray}
R^{(j)}_1 & := & \begin{pmatrix} I_a & -C_j \end{pmatrix},  \nonumber \\
R^{(j)}_2 & := & \begin{pmatrix} 0 & I_b\end{pmatrix},
\end{eqnarray}
\begin{equation}
R^{(j)}:=\begin{pmatrix}  R^{(j)}_1 \\ R^{(j)}_2 \end{pmatrix},
\end{equation}
\begin{equation}
\omega^{(j-1)} := R_1^{(j)} \omega^{(j)}.
\end{equation}
Also define $u^{(s)}=0$, $u^{(s-1)}:=2^s z^{(s-1)}$, and for $j=1,2,\ldots,s-1$ define
\begin{equation}
u^{(j-1)}:=2^{j} z^{(j-1)} + R^{(j)}u^{(j)}.
\end{equation}
Note that $C_j$ and $I_b$ have an equal number of columns.

Decoding is as follows. Starting with  $Q_s x^{(s)} = \omega^{(s)}$ pre-multiply both sides by $R^{(s)}$  apply (\ref{eqn:Qiter-rep}) and the above definitions, in order to get 
for $j=s,s-1,\ldots,1$,
\begin{eqnarray}
y^{(j-1)} + C_j^{tr} z^{(j-1)} - R^{(j)}_2u^{(j)}& =  &  R^{(j)}_2 \omega^{(j)}, \\
Q_{j-1} x^{(j-1)}-u^{(j-1)} & = & \omega^{(j-1)}.
\end{eqnarray}

Note that all entries in $u^{(j-1)}$ are divisible by $2^j$. We start uncovering the bits with
\begin{equation}
x^{(0)} -u^{(0)}  =   \omega^{(0)}
\end{equation}
which is solved for $x^{(0)}$ by
 \begin{equation}
 x^{(0)}= \left( \omega^{(0)}\right)_2
 \end{equation}
 where $\left( x \right)_2$ is the residue of $x$ modulo 2. Successive bits are then recovered by solving, for $j=0,1,\ldots,s-1$, 
\begin{eqnarray}
u^{(j)} & = & Q_j x^{(j)}-\omega^{(j)} \\
z^{(j)} & = & \left( \frac{u^{(j)}}{2^{j+1}}\right)_2 \\
y^{(j)} & = & \left( R_2^{(j+1)} \omega^{(j+1)} - C_{j+1}^{tr}z^{(j)}\right)_2 \\
x^{(j+1)} & = & (x^{(j)}, y^{(j)},z^{(j)}).
\end{eqnarray}
This completes the decoding process.

\section{Achievable Query Ratios}
\label{sec-achievable}
Let $(m_{s-1},n_{s-1})$ be the size of $Q_{s-1}(r)$ and let $n_c$ denote the number of columns of $C_s$ in (\ref{eqn:Qiter}). Note that
$m_{s-1}$ and $n_{s-1}$ grow with $r$.
\begin{theorem}
For $r$ suitably large, $\rho_s(r)$, the  query ratio for $Q_s(r)$ is arbitrarily close to $1/(s+1)$. \end{theorem}
\begin{IEEEproof}
We proceed by induction. The hypothesis has been proved for $s=1$. Assume that it is
true for $Q_{s-1}(r)$ and let $\rho_{s-1}(r)$ be its query ratio.
From~\cite{GrahamSloane:1980}
we know that there exists a constant weight code with weight $2^s$ and minimum Hamming distance $2(2^s-1)$ with $m_{s-1}$ codewords provided $n_c > \sqrt{(2^s)!m_{s-1}}$ and $m_{s-1}$ is suitably large. Thus
$m_s=m_{s-1}+n_c = n_{s-1}\rho_{s-1}(r) + \sqrt{(2^s)! m_{s-1}}$, $n_s=n_{s-1}(1+\rho_{s-1}(r))+\sqrt{(2^s)! m_{s-1}}$ and $\lim_{r \rightarrow \infty} \rho_s(r) =\lim_{r \rightarrow \infty}\frac{\rho_{s-1}(r)}{1+\rho_{s-1}(r)} =\frac{1}{s+1}$.
\end{IEEEproof}

Thus we can achieve an arbitrarily small query ratio, by choosing $s$ and $r$ sufficiently large.

\section{Example Constructions}
\label{sec-examples}
We construct a level-$1$, level-$2$  and level-$3$ query matrices in the examples presented below.
\begin{example}
Let $r=4$. Thus $n=16$ and  $m=10$. We construct $Q_1(4)$ by picking as the rows of $C_1$
the six binary $4$-tuples of weight $2$. Thus
$$
C_1=\begin{pmatrix}
1100\\1010\\1001\\0110\\0101\\0011
\end{pmatrix}
 \implies 
E_1= \begin{pmatrix} 
011110\\101101\\110011\\110011\\101101\\011110
\end{pmatrix}
$$
and $Q_1(4)$ is completely specified. $Q_1(4)$ has query ratio $5/8$.
\label{ex:Q1}
\end{example}

\begin{example}
We construct $Q_2(9)$ by first constructing $Q_1(9)$ as described. $Q_1(9)$ is a $(45,81)$
binary matrix. For $Q_2(9)$, we selected $45$ rows of the incidence matrix for the Steiner system $S(2,4,25)$ tabulated
at \cite{ccrwest:2012}. Note that $|S(2,4,25)|=50= \frac{{25 \choose 2}}{{4 \choose 2}}$.
The matrix $Q_2(9)$ has size $(70,151)$, thus achieving a query ratio of $70/151 < 1/2$.
A decoding rule was implemented for this design and error free decoding was observed 
in a simulation consisting of $10,000$ test vectors.
\label{ex:Q2}
\end{example}

\begin{example}
We construct $Q_3(9)$ by selecting for $C_3$, $70$ rows of the incidence matrix for the Steiner system $S(2,8,64)$~\cite{ccrwest:2012}, which has $|S(2,8,64)|=72$.
The query ratio $\rho$ is larger than in the previous example because the size is not large enough. Unfortunately, there are no larger published $S(2,8,*)$ designs currently available.
\label{ex:Q3}
\end{example}
We close with a graph showing the various results from the paper. The curve labeled LY is the existence result (\ref{eqn:LVbound}).  The curve labeled `Packing' is the result of Thm.~\ref{thm:packing}. The curves labeled $Q_i:GS$ use the bound (\ref{eqn:GSbound}) to estimate the query matrix size, `$Q_3:W$' uses Wilson's theorem (\ref{eqn:Wbound}) to estimate
the size of $C_3$ and the data points are for Examples \ref{ex:Q1}--\ref{ex:Q3}. 

Wilson's theorem guarantees existence for suitably large designs, so this curve needs to be interpreted carefully.
The reason this bound was included was that it more closely matches data for small block designs as can be seen by
how close it comes to the performance of $Q_3(9)$.

Data on larger designs is not available unfortunately.

\begin{figure}[htbp] 
   \centering
   \includegraphics[width=4in]{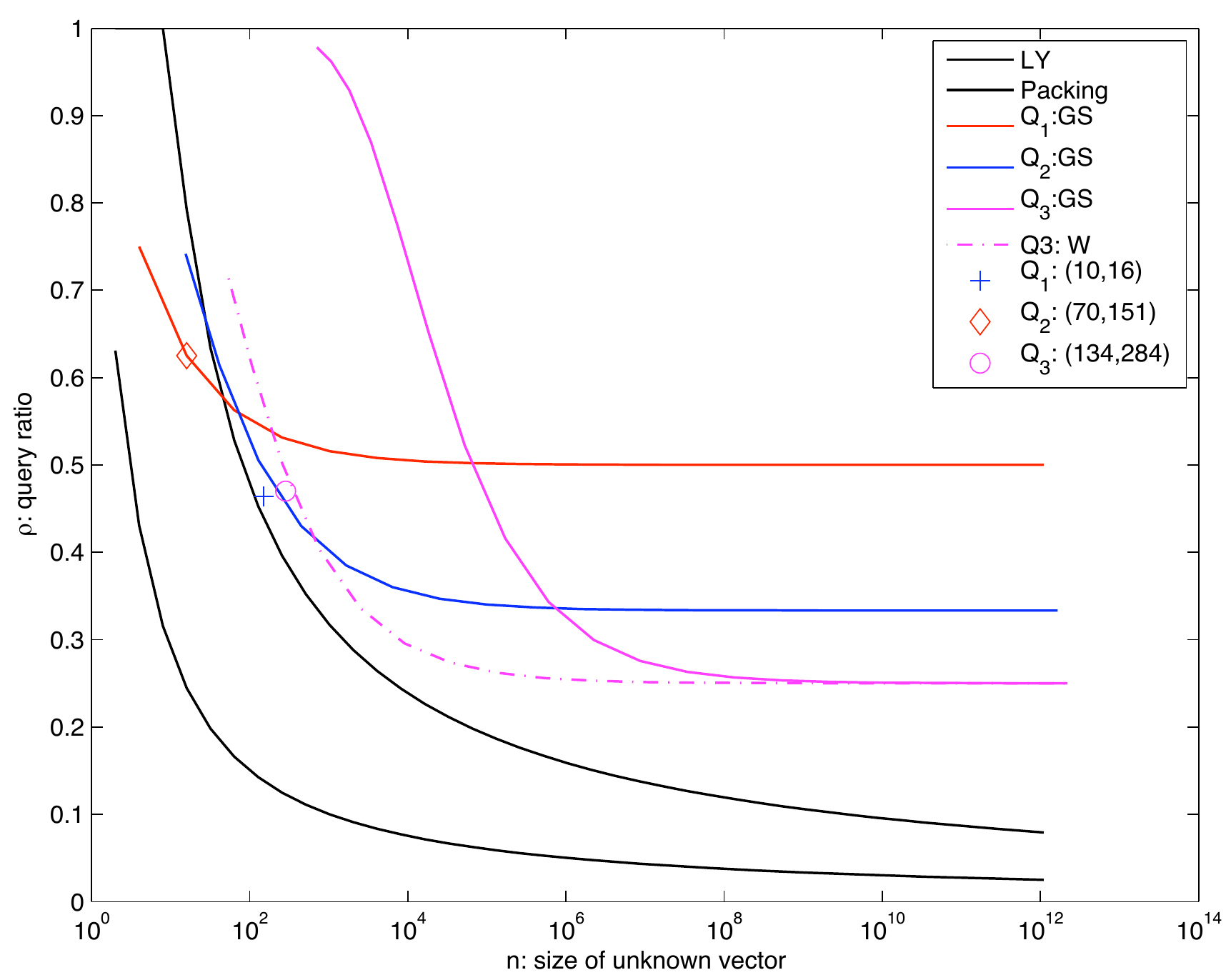} 
   \caption{Query ratio as a function of the size of the unknown vector. Shown are various bounds, performance of $Q_1$, $Q_2$ and $Q_3$ and the constructions of
   Examples~\ref{ex:Q1}--\ref{ex:Q3}.}
   \label{fig:example}
\end{figure}

\section{Conclusions}
\label{sec-conclusions}
Binary query matrices are constructed for the Hamming oracle. 
The construction is algebraic and uses previously known codes of constant Hamming weight with a specified
minimum distance. Starting from a level-1 construction, a sequence of query matrices is constructed
by iterating a simple design rule. Thus a level-$i$ query matrix is constructed using a level-$(i-1)$ matrix, $i>1$.
Our query matrices  are shown to be uniquely identifying,
i.e., it is possible to uniquely determine any unknown binary vector $x$
using the query vectors in a query matrix.
We also establish a connection between our problem and the distinct subset sum problem studied in
the combinatorics literature. To be specific our construction makes use of the
set $\{1,2,4,...,2^n\}$, which is the simplest example of a  set with distinct subset sums.
It is not clear whether the construction presented here can take advantage of other DSS sets presented in
\cite{Bohman:1996}, or whether there is a significant advantage in doing so.

\bibliographystyle{IEEEtran}

\end{document}